\documentclass[aip,showpacs,preprint]{revtex4-1}

\usepackage{amsmath,amssymb,amsthm}

\usepackage{hyperref}
\usepackage{amsthm}
\usepackage{graphicx}
\usepackage{subfigure}
\usepackage{color}
\usepackage{bbm}
\usepackage[normalem]{ulem}
\usepackage{bbm}

\renewcommand{\P}{\mathbb{P}}
\graphicspath{{./Figures/}}

\theoremstyle{plain}
\newtheorem{theorem}{Theorem}

\theoremstyle{definition}

\begin{document}

\title{Optimal system size for complex dynamics in random neural networks near criticality}

\author{Gilles Wainrib}%
\affiliation{Laboratoire Analyse G\'eom\'etrie et Applications, Universit\'e Paris XIII, France. Email: wainrib@math.univ-paris13.fr}
\author{Luis Carlos Garc\'ia del Molino}
\affiliation{Institute Jacques Monod, Universit\'e Paris VII, France. Email: garciadelmolino@ijm.univ-paris-diderot.fr}

\date{\today}%

\begin{abstract}
In this article, we consider a model of dynamical agents coupled through a random connectivity matrix, as introduced in [Sompolinsky et. al, 1988] in the context of random neural networks. When system size is infinite, it is known increasing the disorder parameter induces a phase transition leading to chaotic dynamics. We observe and investigate here a novel phenomenon in the subcritical regime for finite size systems : the probability of observing complex dynamics is maximal for an intermediate system size when the disorder is close enough to criticality. We give a more general explanation of this type of system size resonance in the framework of extreme values theory for eigenvalues of random matrices.
	\end{abstract}

\maketitle

{\bf The human brain is composed of approximately one hundred billion neurons, and each neuron is typically connected through ten thousand synaptic connections. Understanding the dynamical behavior of such complex systems is a key challenge in neuroscience and beyond. The study of mathematical models of many interacting dynamical units appears to be an efficient approach to gain new insights into this problem. Due to the enormous amount of unknown parameters in these models, for instance the synaptic connections strengths between each pair of neurons, a classical modeling strategy is to consider these parameters to be randomly drawn according to a given distribution, giving rise to  \emph{disordered dynamical systems}. In this framework, it is possible to study how the statistical properties of the disorder will affect the dynamical behavior of the system. In the context of neural network, it is known that, at the limit of infinite number of neurons, increasing the level of disorder induces to a phase transition leading to complex chaotic dynamics. Interestingly, the critical regime displays many important features in terms of information processing capacity, and has found applications in the field of machine learning. Although the behavior is rather clear in the limit of infinite system size, the situation is substantially different for finite size systems. In particular, various attractors (equilibria, limit cycles, chaos) may co-exist and contribute to a much richer dynamical repertoire, even in the subcritical regime. We investigate in this article the probability of observing complex dynamics, and report a novel phenomenon of system size resonance: close to criticality, there exists an intermediate system size for which the probability of chaos is maximal. We provide a theoretical explanation of this effect in the framework of random matrix theory. This property may have significant implications for the understanding of dynamical systems of complex networks with a modular structure.}

   \section{Introduction}
   \label{sec:intro}

Most complex systems involve a large number of interacting elements and are often modeled and studied within the framework dynamical systems on random networks. Typical examples range from condensed matter physics  \cite{spinglass}, to biology \cite{sompolinsky,luo}, and social sciences \cite{scott}. In particular,
the hundreds of billions of heterogeneous synaptic connections between neurons in the brain constitute a paradigmatic example of a disordered system of dynamical agents coupled through a random connectivity matrix. To understand the impact of this heterogeneity on brain dynamics, a random neural network model has been introduced in \cite{amari,sompolinsky}: 
\begin{equation}
\mbox{for }1\leq i\leq n,\ \frac{dx_i}{dt} = -x_i + \sum_{i=1}^n J_{ij}\phi(x_j)
\label{eq1}
\end{equation}
where $\mathbf{J}=(J_{ij})$ is a Ginibre \cite{ginibre} random matrix with Gaussian i.i.d entries such that $\mathbf{E}[J_{ij}]=0$ and $\mathbf{E}[J_{ij}^2]=\sigma^2/n$, and where $\phi(\cdot)$ is a smooth odd sigmoid function with unit slope at the origin $\phi'(0)=1$. The variables $x_i$ represent the activity of a neuron $i$ (or neuronal population) and $J_{ij}$ the synaptic weight from neuron $j$ to neuron $i$. Notice that zero is a trivial equilibrium of \eqref{eq1}.

It is known since \cite{sompolinsky} (see also \cite{benarous,toubprl}) that in the limit $n\to \infty$ one can derive a mean field description of the above model and an analysis of the mean field equations \cite{sompolinsky} leads to the two following regimes :

(i) if $\sigma<1$ the only equilibrium point is zero and attracts all trajectories ;

(ii) if $\sigma>1$ there exist infinitely many equilibria and limit cycles but none of them is stable. The only {stable attractors} are chaotic in the sense of positive Lyapunov exponents.

This abrupt transition is also present in terms of linear stability analysis and can be recovered using a classical result of random matrix theory. Indeed, the circular law \cite{girko, tao} states that the empirical distribution of the complex eigenvalues for $n\times n$ random matrices  with i.i.d centered coefficients of variance $n^{-1}$, converges to the uniform measure on the unit disk when $n\to \infty$. Therefore, in the limit $n\to \infty$, the trivial equilibrium zero becomes unstable as $\sigma$ crosses $1$.

Based on a combination of experimental and theoretical considerations, it has been argued in \cite{beggs2008criticality, kitzbichler2009broadband} that living neuronal networks, as well as many other complex systems \cite{bak1996nature}, may evolve in the vicinity of the phase transition also called \textit{the edge of chaos} \cite{langton1990}. Recently, this specific regime has been shown to be particularly important in the field of reservoir computing \cite{jaeger}, a popular class of machine learning algorithms, because of the increased information processing capacity near criticality \cite{criticalreservoir}. 

Therefore, it is of interest to study the behavior of this system close to the phase transition. Although the situation is rather clear in the infinite system size limit, as stressed in \cite{sompolinsky}, the picture is different for finite size systems where, close to criticality, several stable equilibria, limit cycles and chaotic attractors may co-exist. A fine knowledge of the statistical properties of the eigenvalues distribution of random matrices is fundamental to characterize the behavior of various disordered systems, in particular in terms of stability properties \cite{may72}. In the critical region the extreme eigenvalues of $\mathbf{J}$ play a crucial role in the stability in the finite size system, suggesting that its behavior will not be properly described by the mean-field equations. Indeed, in the case of finite size matrices, despite most eigenvalues being inside the unit circle, there is a nonzero probability that eigenvalues close to the border have a modulus slightly larger than 1, leading to unstable modes from the stability point of view.

Our aim is to study the probability of observing spontaneous activity (limit cycle or chaotic attractor) in system \eqref{eq1} near the phase transition, specifically for values of disorder $\sigma$ slightly below criticality. As $n\to \infty$, this probability shall converge to zero, but as we will show, this convergence is not monotonous, this probability is actually reaching a maximum for an intermediate value of $n$. We will first investigate this question directly on system \eqref{eq1} numerically (\textbf{section II}), and then provide a deeper theoretical explanation of this phenomenon of system size resonance through the study of extreme eigenvalues of random matrices (\textbf{section III}). Finally, in \textbf{section IV.} we discuss our results.

  \section{Probability of complex dynamics}
 \label{sec:numerical}


  To determine wether the neural network is activated or not we compute the maximal Lyapunov exponent $\Lambda$ of the attractors. Depending on the value of $\Lambda$ one can distinguish three cases:  (i) if $\Lambda<0$ the attractor is a fixed point (might be other than 0), the network is not activated, (ii) if $\Lambda=0$ the attractor is a limit cycle, the network is activated and exhibits regular activity, and (iii) if $\Lambda>0$ the attractor is chaotic, the network is activated and exhibits irregular activity. In terms of $\Lambda$, the probability of observing spontaneous activity, whether it is chaotic or periodic, is given by $\P[\Lambda\geq0]$.

 In the numerical simulations, an estimation of the maximal Lyapunov exponent of the attractors $\bar{\Lambda}$ is computed using the variational equations as shown in \cite{skokos}. Starting from a random initial condition, we evolve a reference solution $(x_i(t))_{1\leq i\leq n}$ according to \eqref{eq1} and a normalized difference vector on the tangent plane $(\delta(x,t)_i)_{1\leq i\leq n}$ driven by the Jacobian of \eqref{eq1}. At each time step the modulus of $\delta$ is normalized to a small quantity $\delta_0$ so that $(x(t)_i+\delta(x,t)_i)_{1\leq i\leq n}$ is always in the vicinity of the reference solution. After some iterations the reference orbit will be close to one of the attractors of the system and the difference vector will point in the direction corresponding to the maximal Lyapunov exponent of the reference trajectory. Then, we estimate $\bar{\Lambda}=\frac{dt}{T}\sum_{i=1}^{T/dt}\log\frac{|\delta|}{\delta_0}$ where $T$ is the sampling time and $dt$ is the time step. To circumvent 
approximations error in the case of periodic attractors, we set $\bar{\Lambda}=0$ when periodicity is detected directly on the trajectories \footnote{The precision of this method is limited by two reasons asides from truncation error. First, the precision increases as $T/dt$ increases but for obvious reasons we have to keep this quantity finite. Second, the method computes the maximal Lyapunov exponent of the reference trajectory, not the one of the attractor itself. In general we assume that after evolving the system for a while the reference trajectory is close enough to the attractor. This assumption is easy to check by direct examination of the trajectories when the attractor is a limit cycle or a fixed point but it is impossible to distinguish a chaotic trajectory from a long transient regime. However, as we are only interested in the sign of $\Lambda$ the error is irrelevant when $\bar{\Lambda}$ is clearly positive or negative. Only when $\bar{\Lambda} \approx 0$ the error becomes relevant to the 
analysis. Therefore, to detect periodicity, we rely instead on a direct analysis of the trajectories, and we set $\bar{\Lambda}=0$ when periodicity is detected.}
.

\begin{figure}[ht!]
\begin{center}
\subfigure[Probability of spontaneous activity as a function of $n$]{\includegraphics[scale=0.3]{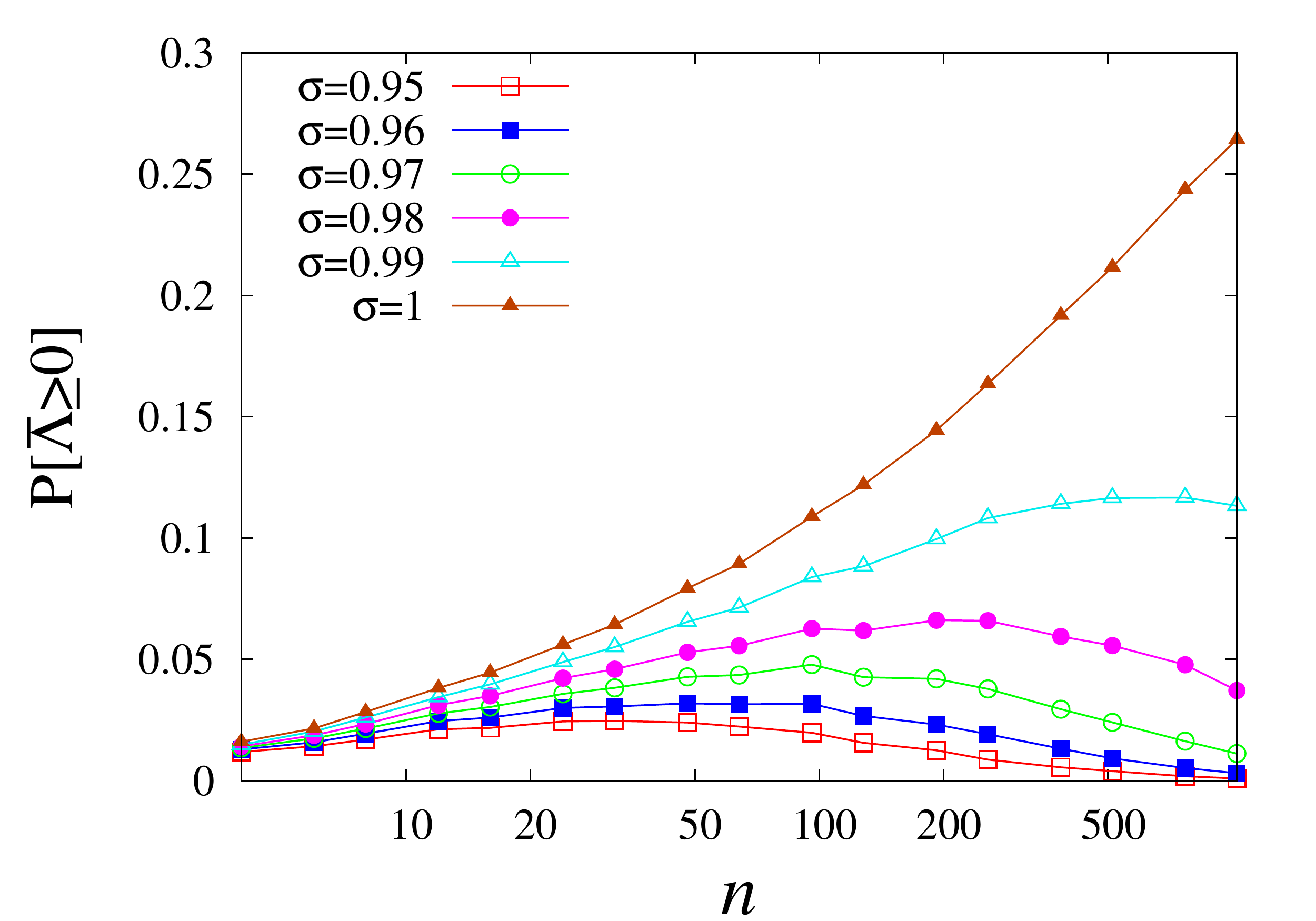}}
\subfigure[Probability of limit cycle as a function of $n$]{\includegraphics[scale=0.3]{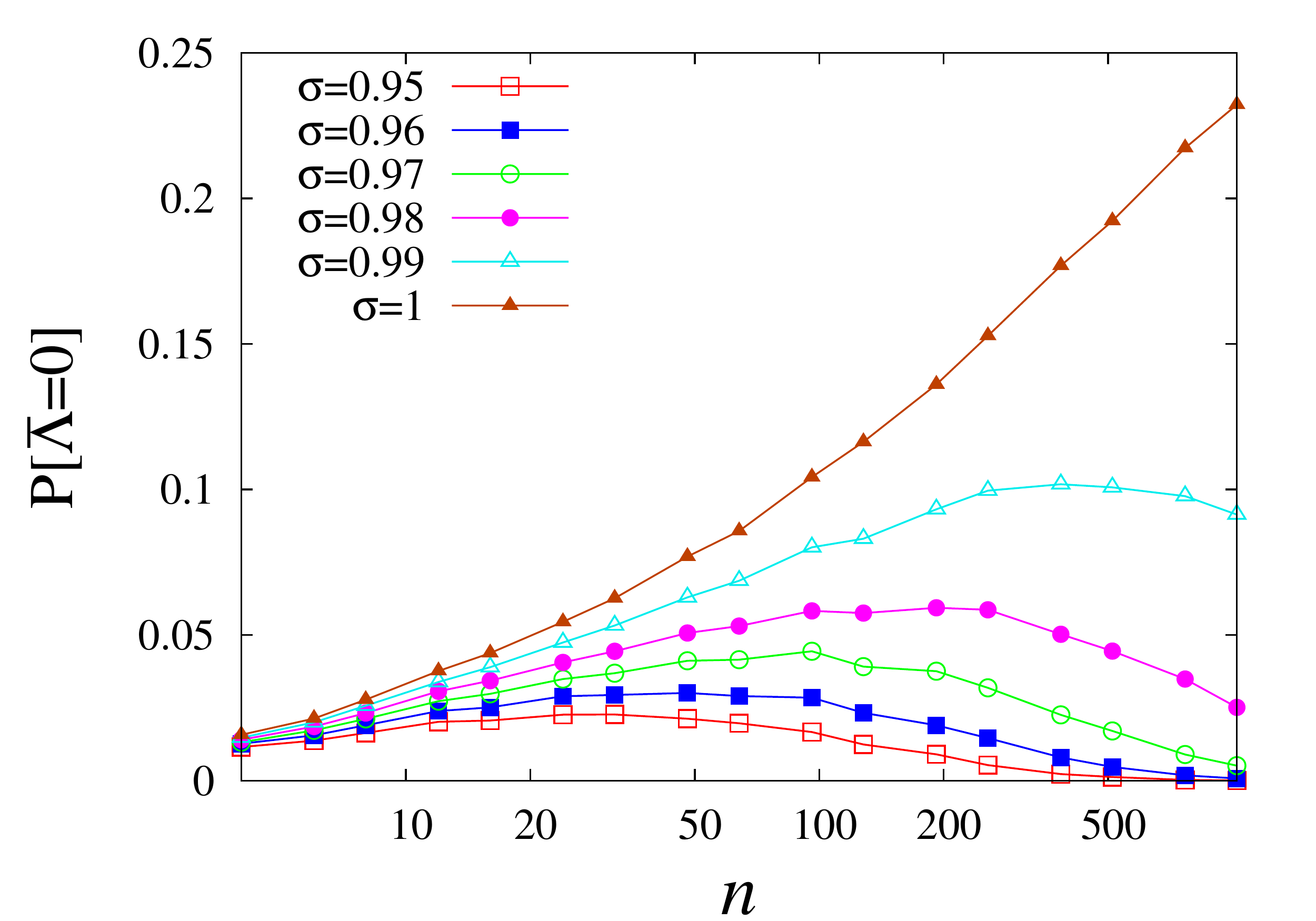}}\\
\subfigure[Probability of chaos as a function of $n$]{\includegraphics[scale=0.3]{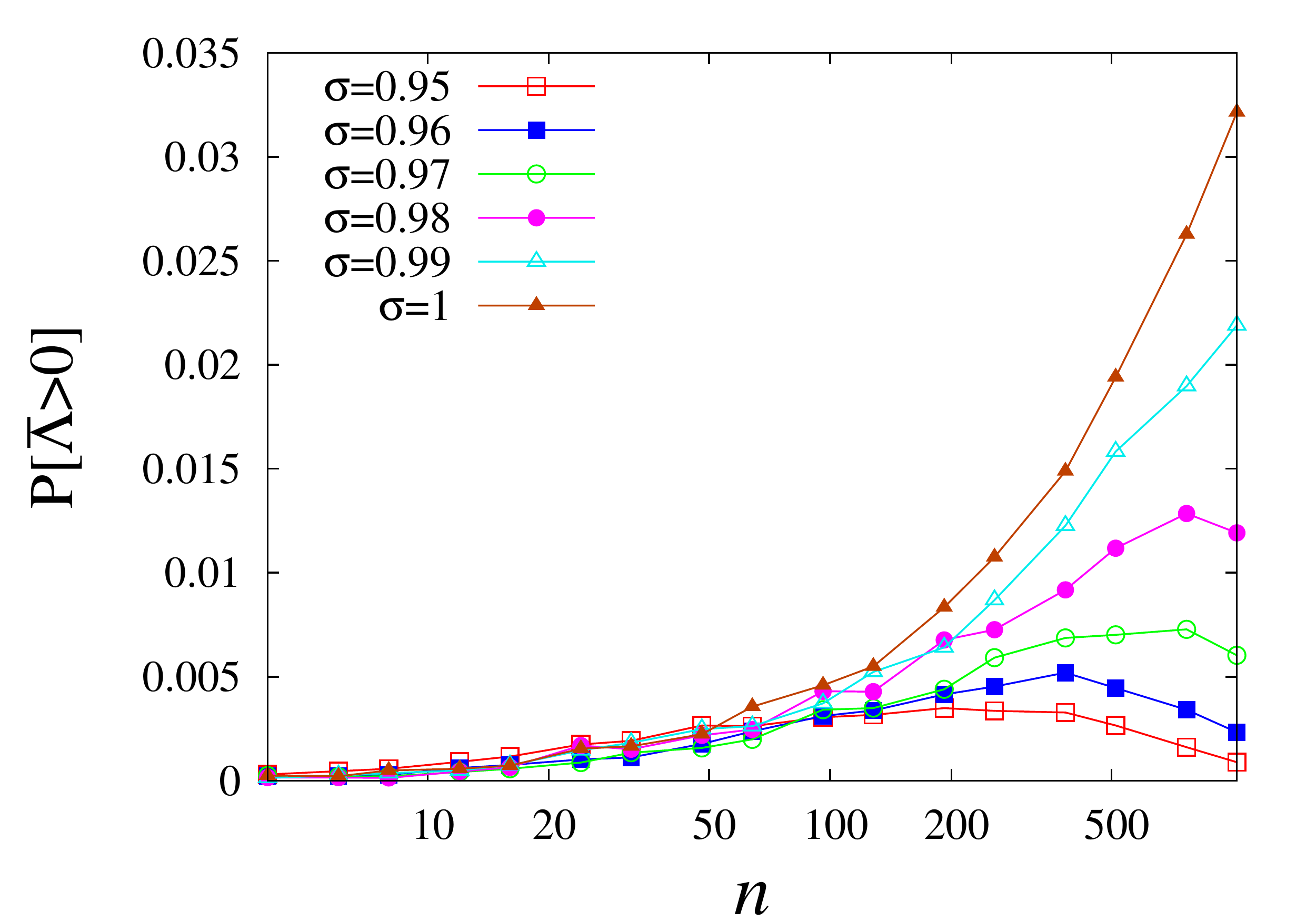}}
\subfigure[Probability of chaos given that spontaneous activity occurs, as a function of $n$]{\includegraphics[scale=0.3]{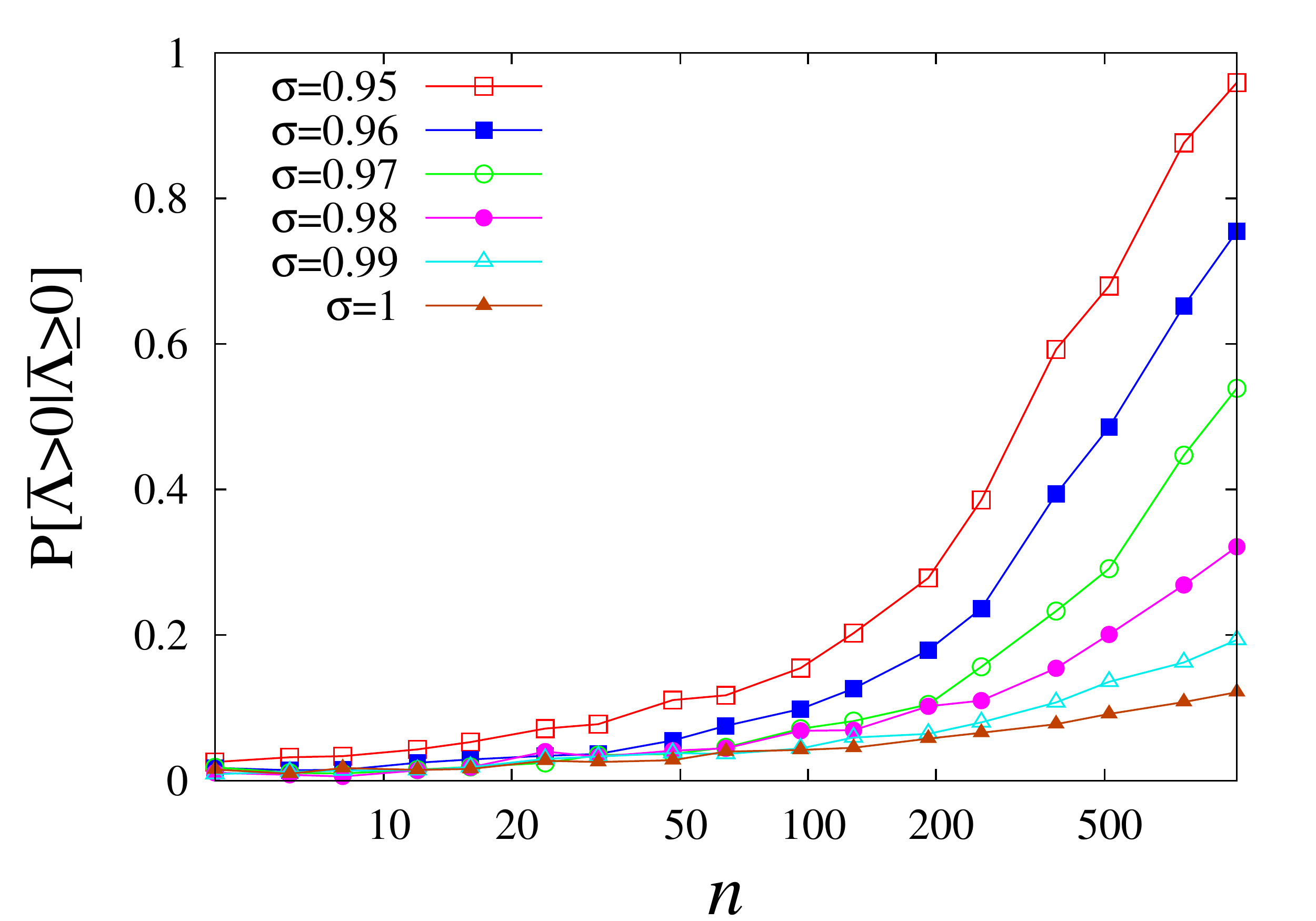}}
\end{center}
\caption{Numerical estimation of the probability of observing (a) spontaneous activity (b) limit cycles  (c) chaotic trajectories and (d) chaotic attractor given spontaneous activity, as a function of $n$ and for different values of $\sigma \in \{0.95,0.96,0.97,0.98,0.99,1\}$. For each matrix size and each value of $\sigma\leq1$, $5.10^4$ realizations of the random matrix have been analyzed and the sigmoid function used was $\tanh(\cdot)$. Standard deviations errors are smaller than the points, therefore not shown.  Numerical estimations of the maximal Lyapunov exponent were performed after cutting a first period of transient dynamics.}
\label{fig:proba}
\end{figure}

   Numerical results are shown in Fig.\ \ref{fig:proba}. The main unexpected phenomenon is that  the probability of observing spontaneous activity for $\sigma<1$, close enough to $1$, displays a maximum for an intermediate value of the network size $n$ (Fig.\ \ref{fig:proba} a.). Surprisingly, this probability first increases with $n$ until an optimal system size $\bar{n}(\sigma)$, and then decreases to zero as expected from \cite{sompolinsky}. Moreover, as $\sigma$ gets closer to $1$, both the optimal size $\bar{n}(\sigma)$  and the probability of spontaneous activity at $n=\bar{n}(\sigma)$ increase. In the limit case $\sigma=1$ we do not observe a maximum and $\P[\bar{\Lambda}\geq0]$ tends to 1 as $n$ becomes larger. 

 The estimations of the probabilities corresponding to limit cycles $\P[\bar{\Lambda}=0]$ (Fig.\ \ref{fig:proba} b.) and chaotic oscillations $\P[\bar{\Lambda}>0]$ (Fig.\ \ref{fig:proba} c.) have a similar behavior. The main difference is that the maxima for $\P[\bar{\Lambda}>0]$ are reached at larger values of $n$  than the ones for $\P[\bar{\Lambda}=0]$. As shown in Fig. \ref{fig:proba} d., this difference means that the larger the system is, the more likely is that the spontaneous activity takes the form of chaotic attractors.
 
Therefore, numerical simulations have revealed a novel type of system size resonance effect, showing an enhancement of the probability of observing complex dynamics for an intermediate system size. In the following section, we study this new phenomenon within the framework of random matrix theory.

   \section{Extreme eigenvalues of random matrices}
   \label{sec:extreme}
   From numerical simulations, we have shown the existence of an intermediate system size which maximizes the probability of observing complex dynamics. This mysterious phenomenon has not been described previously and we intend to provide a theoretical explanation through the lens of random matrix theory. Indeed, as mentioned in \cite{sompolinsky}, the phase transition can be inferred from the study of the linearized problem around the trivial equilibrium: in the limit $n=\infty$, linear stability is lost when $\sigma>1$ as a direct consequence of the circular law. Moreover,  if $(\lambda_k)_{1\leq k\leq n}$ denote the complex eigenvalues of $\mathbf{J}$, and if one defines 
   \begin{equation}
   \rho_1=\max_{1\leq k\leq n} Re(\lambda_k),
   \end{equation}
    then  $\rho_1<1$ implies that the trivial equilibrium is globally attractive. Therefore, a natural quantity to study is the probability $p_n$ that the trivial equilibrium zero is linearly stable. Such a question has a direct translation in terms of random matrix theory:
\begin{equation}
\label{eq:pn}
p_n = \mathbf{P}\left[\rho_1 < 1 \right] .
\end{equation}

First, let us remark the fact that if $\sigma<1$ then $p_n \to 1$ when $n\to \infty$.
We will show below that this probability has a minimum for an intermediate matrix size, which increases to $+\infty$ when $\sigma\to 1^-$. In terms of the original system \eqref{eq1}, this relates to the fact that the probability of observing a spontaneous oscillations or complex dynamics when $\sigma$ is close to $1^-$ displays a maximum for an intermediate system size. 
 The probability $p_n$ is not exactly equal to $\P[\Lambda<0]$, but one only has the inequality $p_n\leq\P[\Lambda<0]$. Indeed, it is possible to have $\rho_1>1$ and still $\Lambda<0$ due to the existence non trivial attracting equilibrium points.  However, we expect both quantities to have similar properties.       
    
    To study $\rho_1$, we need to build upon the extreme value theory for the eigenvalues random matrices. Contrary to classical extreme value theory, there is a strong dependance structure among the eigenvalues, which is well described within the framework of determinental point processes. In fact, the theory of extremal eigenvalues of complex Gaussian matrices has been investigated recently in \cite{Bender}.

\begin{figure}[ht!]
\center
\includegraphics[scale=0.3]{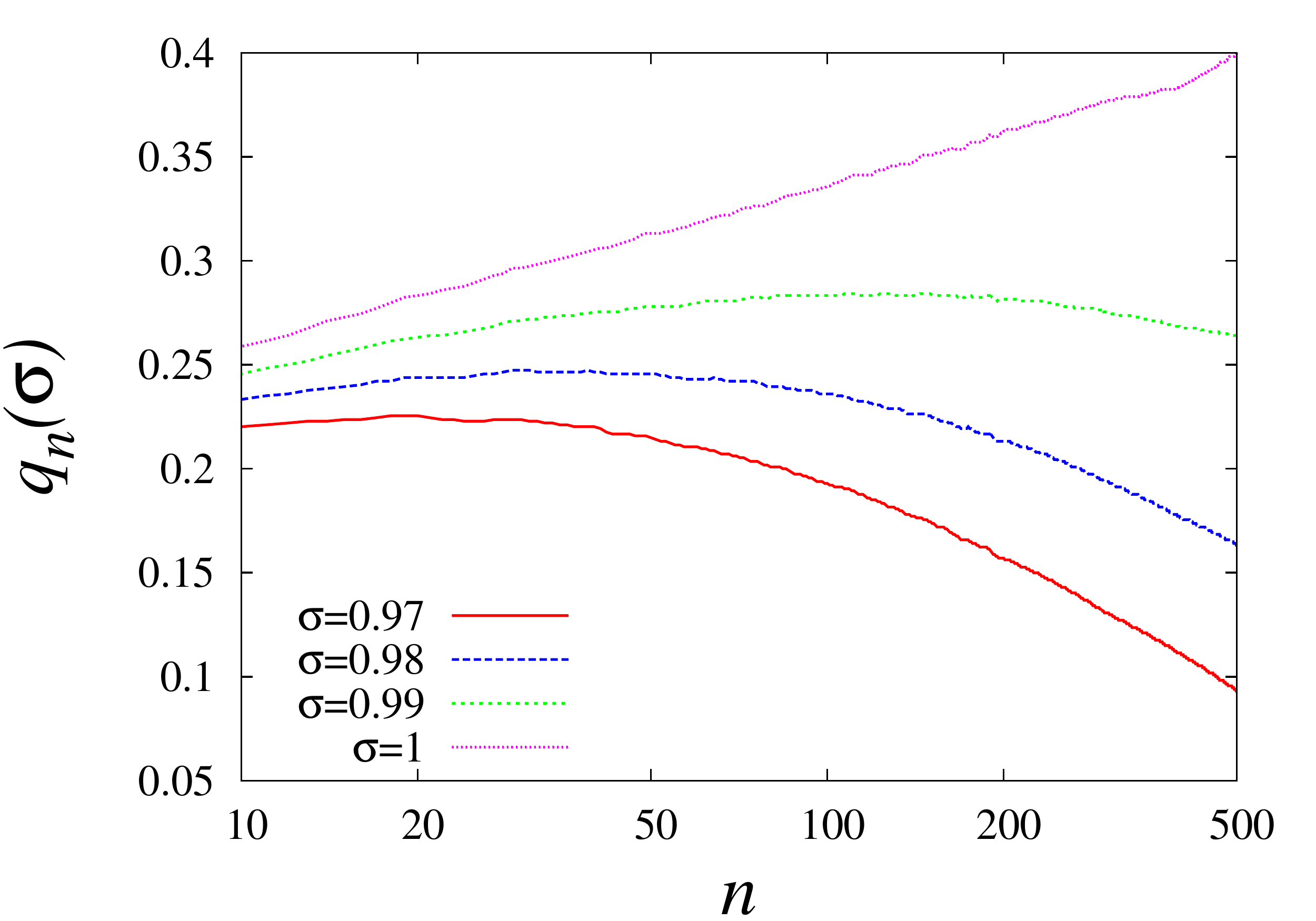}
\caption{Real Ginibre Ensemble : Numerical estimation of  $q_n(\sigma)$ as a function of $n$, for different values of $\sigma = 0.97, 0.98, 0.99, 1$ (from bottom to top line). A total of $5.10^5$ Monte-Carlo runs were computed for each different matrix size and value of $\sigma$.}
\label{fig:landscape}
\end{figure}

    Based on these results, we explain the optimal size phenomenon in the case of complex matrices. In the case of real Gaussian matrices, the situation is more delicate and we only provide numerical results in Fig.\ \ref{fig:landscape} showing the same phenomenology.

More precisely, for $\sigma >0$, we consider $\{\lambda_k^{(n)}(\sigma)\}_{1\leq k\leq n}$ the complex eigenvalues of a $n \times n$ random matrix $\mathbf{J}$ such that the coefficients $J_{ij}$ are i.i.d complex Gaussian random variables on $\mathbb{C}\equiv \mathbb{R}^2$ with $J_{ij} \approx \mathcal{N}(0,\frac{\sigma^2}{2n}I_2)$.

\noindent We denote the maximum of the real parts of the eigenvalues of $\mathbf{J}$ by:
\begin{equation}
\lambda^{(n)}_{max}(\sigma):=\max_{1\leq k\leq n} Re(\lambda_k)
\end{equation}
We are interested in the probability that all the eigenvalues have a real part smaller than $1$:
\begin{equation}
p_n(\sigma) := \mathbb{P}\left[\lambda^{(n)}_{max}(\sigma) < 1 \right]
\end{equation}
and its complementary probability $q_n(\sigma)=1-p_n(\sigma)$, that is the probability that there exists at least one eigenvalue with a real part larger than $1$. Our main theoretical result is the following:

\begin{theorem} For any $\sigma \in (0,1)$, the integer $n^*(\sigma)$ such that $q_{n^*(\sigma)}= \displaystyle{\sup_{n\in\mathbb{N}}}\  q_n(\sigma)$ satisfies:
\begin{equation}
\displaystyle{\lim_{\sigma\to 1^-}} n^*(\sigma) = +\infty 
\label{eq:thm2.1}
\end{equation}
Moreover, $q_n(\sigma)$ can be made arbitrary close to $1$:
\begin{equation}
\displaystyle{\lim_{\sigma\to 1^-}} q_{n^*(\sigma)} =1 
\label{eq:limit}
\end{equation}

\end{theorem}
Heuristically, this property stems from a competition between (i) the fact that when the number of eigenvalues $n$ increases, the maximal real part tends to be higher, as in classical extreme value theory, and (ii) the convergence of the spectral density to the unit disk that implies a concentration of $\lambda^{(n)}_{max}(\sigma)$ close to $\sigma<1$.


\begin{proof}

The proof of this result is based on \cite{Bender}, where the asymptotic law for large $n$ of the largest real part of eigenvalues for a class non-Hermitian random matrix is studied, developing extreme value theory for determinental processes. Essentially, Theorem 2.5 of \cite{Bender} states that:
\begin{equation}
\displaystyle{\lim_{n\to\infty}} \mathbb{P}\left[\lambda^{(n)}_{max}(1) -1\leq \frac{1}{2\sqrt{n \log n}}t + c_n\right] = e^{-e^{-t}}
\label{bender}
\end{equation}
for any $t\in \mathbb{R}$, with
\begin{equation}
c_n = \frac{1}{2\sqrt{2}}\sqrt{\frac{\ln(n)}{n}}-\frac{\frac{5}{4\sqrt{2}}\ln(\ln(n))-\ln(2^{1/4}\pi)}{\sqrt{n\ln(n)}}
\end{equation}
From this result, we claim that for any $\delta \in (0,1)$:
\begin{equation}
\displaystyle{\lim_{n\to\infty}} \mathbb{P}\left[\lambda_{max}^{(n)}(1)<1+(1-\delta)c_n\right]=0
\label{claim1}
\end{equation}
Indeed, let $\delta\in (0,1)$ and  $\epsilon>0$ fixed, and $t_{\epsilon}<0$ such that $e^{-e^{-t_{\epsilon}}}<\epsilon$. Then from (\ref{bender}) there exists $n_0\geq 1$ such that for all $n\geq n_0$:
\begin{equation}
\mathbb{P}\left[\lambda_{max}^{(n)}(1)<1+c_n\left(1+\frac{t_{\epsilon}}{2c_n\sqrt{n \log n}}\right)\right] <\epsilon
\end{equation}
Moreover, as $c_n\sqrt{n \log n} \to \infty$ when $n\to\infty$, one can choose $n_0$ such that $-\delta < t_{\epsilon}/(2c_n \sqrt{n\log n})$. Then, by monoticity of the distribution function, we deduce:
\begin{eqnarray*}
&\mathbb{P}&\left[\lambda_{max}^{(n)}(1)<1+c_n(1-\delta)\right] \\
&\leq& \mathbb{P}\left[\lambda_{max}^{(n)}(1) < 1+ c_n\left(1+\frac{t_{\epsilon}}{2c_n\sqrt{n \log n}}\right)\right]<\epsilon
\end{eqnarray*}
which proves claim (\ref{claim1}).

\noindent Now we want to show that $p_n(\sigma)$ can be made arbitrarily small when $\sigma \to 1^-$. First,  since the law of $\lambda^{(n)}_{max}(\sigma)$ is the same as the law of $\sigma\lambda^{(n)}_{max}(1)$, we observe that:
\begin{eqnarray*}
&\mathbb{P}&\left[\lambda_{max}^{(n)}(1)<1+(1-\delta)c_n\right] \\
&=& \mathbb{P}\left[\lambda_{max}^{(n)}(\sigma)<1+\sigma(1-\delta)c_n -(1-\sigma)\right]:=f_n
\end{eqnarray*}
We know that $f_n\to 0$ when $n\to \infty$, so there exists $n_1 \geq 1$ (independent of $\sigma$) such that for all $n\geq n_1$, $f_n<\epsilon$. Moreover, since $c_n\to 0$, we can choose $\tilde{n}(\sigma)$ such that for all $n\geq \tilde{n}(\sigma)$, $\sigma(1-\delta)c_n -(1-\sigma)>0$, say we choose it such that $(1-\delta)c_{\tilde{n}^*(\sigma)} \sim 2(1-\sigma)/\sigma$. As $\tilde{n}(\sigma) \to \infty$ when $\sigma \to 1^-$, there exist $\sigma_0\in (0,1)$ such that $\tilde{n}(\sigma_0)\geq n_0$. Then, by monotonicity of the distribution function, we deduce that:
\begin{equation}
p_{\tilde{n}(\sigma_0)}(\sigma_0)\leq f_{\tilde{n}(\sigma_0)} < \epsilon
\end{equation}
As a consequence, we deduce that $n^{*}(\sigma) \to \infty$ when $\sigma \to 1^-$ as stated in \eqref{eq:thm2.1} and that $q_{n^*(\sigma)} \to 1$ as stated in \eqref{eq:limit}.
\end{proof}

As far as the real Ginibre ensemble is concerned (real coefficients), the above theoretical analysis does not apply directly. Indeed, the spectral joint probability density in this case lacks the determinental structure, which is key in the work of \cite{Bender}. However, very recently extreme value theory for the spectral radius of real matrices has been studied \cite{rider}, suggesting that a similar result could be obtained in this context. In fact, numerical simulations confirm that the same phenomenon occurs. In Fig.\ \ref{fig:landscape}, a numerical estimation of $q_n(\sigma)$ is computed and shows that for $\sigma$ close enough to $1$, $q_n(\sigma)$ first increases then decreases as a function of $n$, similarly to the maximum observed on the maximal Lyapunov exponent.
\section{Discussion}

This phenomenon of system size resonance may appear in a wider class of problems, involving the maximum of a set $n$ convergent random variables. It is instructive to discuss this idea on a toy example. Consider a sequence of families of real i.i.d random variables $\left(X^{(n)}_i\right)_{1\leq i\leq n}$ and denote $M_n=\max(X^{(n)}_i;\ 1\leq i\leq n)$. We assume that $\mathbb{P}[X^{(n)}_k\leq x]=a_n=1-b_n$ is increasing to $1$ when $n\to \infty$ for a given $x \in \mathbb{R}$. By the i.i.d assumption, one writes:
   \begin{equation}
   \mathbb{P}[M_n> x]= 1- (1-b_n)^n
   \label{eq:mn}
   \end{equation}
   We deduce that if $b_n=o(n^{-1})$ then $\mathbb{P}[M_n> x] \to 0$ when $n\to \infty$. Now we can wonder whether this convergence is monotonic or if there exists an intermediate value of $n$ for which $\mathbb{P}[M_n > x]$ is maximal. This can be answered by differentiating \eqref{eq:mn} with respect to $n$ (considered as a real variable) and looking $n$ such that:
   \begin{equation}
   a_n \log(a_n) + n a'_n = 0
   \end{equation}
    We find for instance that if $a_n =1- e^{-\kappa n}$ with $\kappa < 1$, then $\mathbb{P}[M_n > x]$ is maximal for $n$ around $\kappa^{-1}$.

   The real parts of the eigenvalues of $\mathbf{J}$ satisfy similar assumptions as the toy model, except the strongest one that is independence. This difficulty has been solved using the approach of determinental processes in \cite{Bender}, which was the starting point of our analysis.
 
 Moreover, one can view the results on the Lyapunov exponent as a consequence of this general principle. There is a competition between two opposite phenomena as the system size increases. On the one hand, consistent with an extreme value behavior, increasing system size increases the likelihood of obtaining a large maximal Lyapunov exponent, since one looks at the maximum on $n$ variables. On the other hand, as $n$ increases, self-averaging principle drives the system towards the mean field behavior which converges to the trivial zero solution for $\sigma< 1$. This competition results in the emergence of an intermediate system size which enhances the probability of observing spontaneous activity.

In terms of perspectives, our result may shed a new light on the behavior of modular networks, composed by clusters of randomly connected agents and sparse connections between clusters. Indeed, each cluster size may be viewed as a parameter that controls the propensity to generate complex dynamics within each cluster.

Finally, although we have carried out the numerical investigation of the maximal Lyapunov exponent only on the random neural network model \eqref{eq1}, the theoretical analysis of $\lambda^{(n)}_{max}$ using random matrix theory is much more general and suggests that the system resonance phenomenon may be observed in various complex systems, and in particular in disordered systems close to the phase transition.


\section*{Acknowledgments} 

G.W. wants to thank Amir Dembo and Ofer Zeitouni for their help in the theoretical part of this paper, and Lenya Ryzhik for his support during year 2010-2011 at Stanford University, where the starting part of this work was done. The authors also want to thank Jonathan Touboul and Khashayar Pakdaman for helpful discussions.

\medskip

\bibliographystyle{aiprev4-1} 
\bibliography{biblio}
\end{document}